\documentclass[
prb,preprint,groupaddress,showpacs,superbib,floats]{revtex4}

\usepackage{graphicx}
\usepackage{amsmath}
\usepackage{times}
\usepackage{verbatim}

\begin{document}

\bibliographystyle{apsrev}



\title{Mesoscopic spin confinement during acoustically induced transport}


\author{J. A. H. Stotz}
\email[]{jstotz@physics.queensu.ca}
\author{P. V. Santos}
\author{R. Hey}
\author{K. H. Ploog}
\affiliation{Paul-Drude-Institut f\"ur Festk\"orperelektronik, Hausvogteiplatz 5--7, 10117 Berlin, Germany}


\date{\today}


\begin{abstract}

Long coherence lifetimes of electron spins transported using moving potential dots are shown to result from the mesoscopic confinement of the spin vector.  The confinement dimensions required for spin control are governed by the characteristic spin-orbit length of the electron spins, which must be larger than the dimensions of the dot potential.  We show that the coherence lifetime of the electron spins is independent of the local carrier densities within each potential dot and that the precession frequency, which is determined by the Dresselhaus contribution to the spin-orbit coupling, can be modified by varying the sample dimensions resulting in predictable changes in the spin-orbit length and, consequently, in the spin coherence lifetime.

\end{abstract}



\maketitle


\def\AlGaAs7{Al$_{0.3}$Ga$_{0.7}$As}

\def\lSAW{\lambda_\mathrm{SAW}}
\def\vSAW{v_\mathrm{SAW}}
\def\fSAW{f_\mathrm{SAW}}
\def\wSAW{\omega_\mathrm{SAW}}
\def\kSAW{k_\mathrm{SAW}}
\def\pSAW{\phi_\mathrm{SAW}} 
\def\PSAW{\Phi_\mathrm{SAW}} 

\def\lDQD{\lambda_\mathrm{DQD}}
\def\vDQD{v_\mathrm{DQD}}
\def\lCAP{\lambda_\mathrm{CAP}}
\def\vCAP{v_\mathrm{CAP}}

\def\Pl{P_\ell}

\def\llaser{\lambda_\mathrm{L}}

\def\pz{$\rho_z$ }
\def\Bint{$\mathbf{B}_\mathrm{int}(\mathbf{k})$}
\def\Bintdqd{$\mathbf{B}_\mathrm{int}(\mathbf{k}_\mathrm{DQD})$}

The ability to store and transport quantum excitations is a critical step towards the application of quantum effects to information processing.  This also pertains to spintronic devices based on semiconductor nanostructures, where  electron spins can be used to represent quantum bits.  Understanding and limiting the various spin decoherence mechanisms has become, therefore, an important field of research that has seen considerable effort.  The high material quality of GaAs-based semiconductors has reduced extrinsic spin scattering to the point where the primary decoherence mechanism for moving spins results from intrinsic D'yakonov-Perel' (DP) spin dephasing.\cite{Dyakonov72}  DP effects typically arise from the random thermal motion of the electron spins in the effective, internal magnetic field \Bint\ associated with the spin-orbit splitting of the conduction band for electrons with non-zero wave vector $\mathbf{k}$.  In the presence of \Bint, the coherence of an initially polarized electron spin ensemble is lost as individual spins follow distinct random walks---each with a different set of non-commutative rotations about \Bint.  DP spin dephasing can be controlled through motional narrowing, whereby rapid momentum scattering times $\tau_\mathrm{p}$ reduce the precession angles during the random walk (and, consequently, increase the spin dephasing time $\tau_\mathrm{s}$) according to the inverse relationship $\tau_\mathrm{s} \sim \tau_\mathrm{p}^{-1}$.\cite{Dyakonov72,Kiselev00}  This motional narrowing mechanism has been invoked to explain the long spin coherence in \emph{n}-type GaAs.\cite{Kikkawa98}

An alternative process to coherently transport spins relies on the use of mobile potentials with mesoscopic, micron-sized dimensions.  In fact, we have recently demonstrated that DP dephasing can be significantly reduced using mobile confinement potentials induced by coherent acoustic phonons.\cite{Stotz05b}  The phonons, generated in the form of surface acoustic waves (SAWs), create a moving, three-dimensional piezoelectric confinement potential (referred to as dynamic quantum dots; DQDs) that coherently transports spin-polarized electrons with the acoustic velocity over long distances (on the order of $100~\mu$m). One interesting question, which will be the subject discussed here, regards the mechanisms leading to the reduced DP dephasing.  Two possibilities were originally proposed.\cite{Stotz05b}  The first suggests that the spin lifetime enhancements arise from motional narrowing associated with the high local electron density within the DQDs, similar to the effects observed in GaAs quantum wells (QWs).\cite{Srinivas93,Sandhu01}  The second possibility lends itself to the fact that when spins are mesoscopically confined to dimensions smaller than the spin-orbit length $\lambda_\mathrm{SO}$, defined as the ballistic transport distance required for a precession angle of 1~rad around \Bint, random spin precession due to thermal motion becomes suppressed, and the DP spin dephasing is limited.\cite{Kiselev00, Malshukov96, Zumbuhl02, Zaitsev05, ZumbuhlPhD04, CHChang04, Holleitner06}  In this case, the coherence enhancement intuitively arises from the motional narrowing associated with the electron scattering on the potential boundaries.

In this Rapid Communication, we unambiguously show that the long spin coherence lengths observed during transport via DQDs result from mesoscopic confinement effects. In fact, spin transport measurements performed by varying the density of optically injected electrons over an order of magnitude demonstrate that the spin coherent transport length, $l_s$, and hence the spin lifetime, is not affected by the local electron concentration.  In contrast, $l_s$ reduces dramatically when the spin-orbit length becomes comparable to or less than the lateral size of the DQD, $L_\mathrm{DQD}$.  We examine this effect through experiments in which $\lambda_\mathrm{SO}$, which is primarily determined by Dresselhaus spin-orbit effects, is varied by changing the thickness of the GaAs QWs and, in particular, show that the experimental results are consistent with $l_s \propto (\lambda_\mathrm{SO})^2$.  The important implications of this confinement, whereby motional narrowing effects do not depend on carrier densities, result in the ability to control spin coherence during transport down to the single spin level.


The DQDs are produced by the interference of two SAW beams propagating along the $\langle 110 \rangle$ surface directions of a GaAs QW sample.\cite{Alsina04, Stotz05}  Three single-QW samples with \AlGaAs7 barriers were grown by molecular-beam epitaxy on GaAs (001) semi-insulating substrates. Two were designed with thicknesses of 12 and 20~nm and placed 390~nm below the surface while the third, 30~nm-thick QW was placed 175~nm below the surface.  The SAWs are excited by applying a radio-frequency signal to two aluminum split-finger interdigitated transducers deposited on the sample surface using standard lithography protocols, and each beam has a linear power density between 2 and 7~W/m.  The SAWs have a wavelength $\lSAW$ of 5.6~$\mu$m, corresponding to a frequency $\Omega_\mathrm{SAW}/2\pi$ of 519~MHz at a sample temperature of 12~K and propagate with a well-defined phase velocity of $\vSAW=2907$~m/s.  The type-II piezoelectric potential generated by the interference of the two plane waves confines and transports the photogenerated electrons and holes within a 120$\times$120~$\mu$m$^2$ array of DQDs, with the diameter of each dot $L_\mathrm{DQD}$ being approximately 1~$\mu$m.  The DQD array propagates along a $\langle 100 \rangle$ direction with a velocity $\vDQD=\sqrt{2}\vSAW$ and has a periodicity $\lDQD=\sqrt{2}\lSAW$.  The measurements were performed at sample temperatures of either 4.2 or 12~K.  As has been previously reported, the electron spin coherence is insensitive to temperatures in this range.\cite{Stotz06}

The coherent spin transport was monitored by microscopic photoluminescence (PL) measurements.\cite{Stotz05b, Stotz06}  A circularly polarized, 768~nm laser beam was focussed onto the sample to photogenerate spin-polarized electrons and holes at a position G.  The carrier densities $n$ are estimated by $n = 2\pi P_{ph}  e^{-\alpha d_\mathrm{QW}} / (E_{ph} \Omega_\mathrm{DQD})$, where $P_{ph}$ is the incident light power, $E_{ph}$ is the photon energy, $\alpha$ is the absorption coefficient in the GaAs QW, $d_\mathrm{QW}$ is the quantum well width, and $\Omega_\mathrm{DQD} = \Omega_\mathrm{SAW}$ is the DQD frequency.  After excitation, the carriers are spatially separated by the piezoelectric potential onto different phases of the DQD lattice and transported along well defined channels.  This efficient charge separation by the acoustically induced potential strongly suppresses spin exchange scattering via the Bir-Aronov-Pikus mechanism during transport.\cite{Bir75,Sogawa01}  While some recombination occurs due to electronic traps in the DQD channel, most luminescence is observed near the edge of a semi-transparent metal strip M that partially screens the piezoelectric potential of the DQDs and allows the electrons and holes to recombine.  The degree of circular polarization  $\rho_z=(I_\mathrm{R}-I_\mathrm{L})/(I_\mathrm{R}+I_\mathrm{L})$ of the luminescence near M can then be measured, where $I_\mathrm{R}$ and $I_\mathrm{L}$ are the right and left circular components of the PL.  The dependence of \pz on the transport distance is mapped by varying the separation between G and M.  Because of the rapid scattering of hole spins in GaAs,\cite{Baylac95} \pz correlates well with the net electron spin population.

%
\begin{figure}[tb]
\centering
\includegraphics[width=7cm, clip=true]{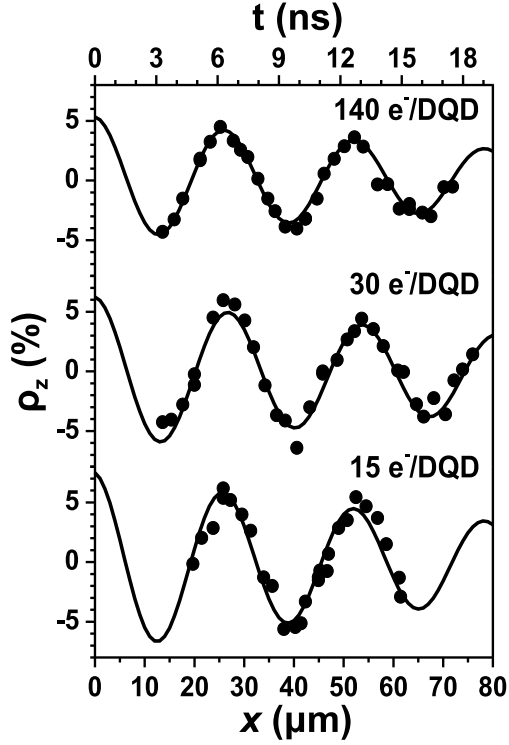}
\caption{Spatial dependence of \pz recorded at varying carrier densities in a 20~nm thick QW. The symbols and solid lines represent the measured values of \pz and the numerical fits, respectively.  All curves provide the spin coherence lengths $l_s$ in the range 110$\pm$30~$\mu$m.  The time axis $t$ is determined by $t=x/\vDQD$.}
\label{excdep}
\end{figure}

Figure \ref{excdep} shows the spatial dependence of \pz for three different electron densities ranging across an order of magnitude (from 15 to 140 electrons per DQD), which correspond to volume (area) concentrations of approximately 10$^{14}$ to 10$^{15}$~cm$^{-3}$ (10$^{8}$ to 10$^{9}$~cm$^{-2}$).  The measured values of polarization were fit with a function of the form $\rho_z(x) = \rho_0 e^{-x/l_s} \cos(\Omega_\mathrm{L}^\mathrm{D}(k_\mathrm{DQD}) x/v_\mathrm{DQD})$, where $\rho_0$ represents the initial spin polarization at G, and $l_s$ is the spin coherence length.  The oscillations in \pz result from the precession of the electron spins around \Bintdqd\ with a frequency $\Omega_\mathrm{L}^\mathrm{D}$ during transport.  The coherent precession observed here occurs in the absence of an external magnetic field and is, for the present sample, primarily related to the \Bintdqd\ associated with the spin-orbit contribution due to the lack of bulk inversion symmetry in the zinc-blende crystal (Dresselhaus term).\cite{Dresselhaus55, Stotz06b}  Consequently, the Larmor frequency of the electron spin precession can be described by
\begin{equation}
\Omega_\mathrm{L}^\mathrm{D}(k_\mathrm{DQD})=\frac{\gamma}{\hbar} k_\mathrm{DQD} \langle k_z \rangle^2=\frac{\gamma}{\hbar} k_\mathrm{DQD} \left(\frac{\pi}{d_{eff}}\right)^2,
\label{larmordqd}
\end{equation}
where $\gamma$ is the spin-orbit parameter, $k_\mathrm{DQD} = m^* v_\mathrm{DQD} / \hbar$ is the average momentum of the electrons within the DQDs, $m^*$ is the electron effective mass, $k_z$ describes the momentum due to the QW confinement, and $d_{eff}$ is the effective QW thickness including the penetration $d_0$ of the electron wavefunction into the \AlGaAs7 barrier layer.\cite{Eppenga88}  The latter was calculated using a tight-binding approach yielding a value of $d_0 = 2.1$~nm for each barrier.  There is also a contribution from the Bychkov-Rashba (BR) term\cite{Bychkov74} related to a structural inversion asymmetry induced, for example, by the vertical component of the piezoelectric field, but both are small for the present experimental conditions and will be neglected.\cite{Stotz06b}

For the electron densities presented in Fig.~\ref{excdep}, the spin coherence lengths $l_s$ are comparable and $\geq 100~\mu$m.  Likewise, the coherence times $T_2^*=l_s/v_\mathrm{DQD}$ of the electron spin microensemble within each DQD remain essentially unchanged.  This is in stark contrast to lifetime measurements on unconfined systems, such as bulk GaAs\cite{Kikkawa98,Dzhioev02} and GaAs QWs\cite{Srinivas93,Sandhu01}, where the spin lifetime has been shown to be strongly carrier dependent.  The long spin lifetimes observed during transport by DQDs cannot, therefore, be attributed to motional narrowing resulting from the mechanisms discussed in previous reports.\cite{Kikkawa98,Dzhioev02,Sandhu01,Leyland06}  Instead, we attribute motional narrowing effects to the DQD piezoelectric confinement of the electron spins.  We argue that the confinement is effective because the size of the DQD $L_\mathrm{DQD}$ is sufficiently small to prevent large precession angles of individual spins during random thermal motion within the DQDs.  The effect of confinement on quantum coherence has been previously studied experimentally\cite{Zumbuhl02} as well as theoretically\cite{Malshukov96,Zaitsev05} in the discussion of weak localization of electrons in a stationary quantum dot with dimensions smaller than the spin-orbit length $\lambda_\mathrm{SO}$.

In the context of the enhanced, long-range transport of quantum states presented here, it is thus anticipated that $\lambda_\mathrm{SO}$ has a larger spatial extent than the approximately $1~\mu$m size of the DQDs ($L_\mathrm{DQD}$).  As mentioned above, the spin-orbit length $\lambda_\mathrm{SO}$ can be intuitively characterized by the distance it takes a spin to precess 1 radian around \Bint.\cite{ZumbuhlPhD04,Kiselev00}  Concerning the contribution to $\lambda_\mathrm{SO}$ due to the Dresselhaus spin-orbit interaction, the temperatures and carrier densities for a QW system relevant to the experiment conditions allow the linear term in $\mathbf{k}$ to dominate over the cubic term.\cite{Eppenga88}  As a result, the Larmor precession frequency associated with the random motion $\Omega_\mathrm{L}^\mathrm{D}(k_\mathrm{F})$ is obtained from Eq.~(\ref{larmordqd}) by replacing of $k_\mathrm{DQD}$ by the Fermi wavevector of the electrons $k_\mathrm{F}$.  As discussed above, the spin-orbit contribution to the Larmor precession from the BR-term and from the induced strain are small compared to the Dresselhaus contribution and will be neglected. Consequently, this approximation results in an isotropic $\lambda_\mathrm{SO}$ given by
\begin{equation}
\lambda_\mathrm{SO} = \frac {v_\mathrm{F}}{\Omega_\mathrm{L}^\mathrm{D}} = \frac{\hbar^2 (d_{eff})^2} {\pi^2 \gamma m^*}.
\label{solength}
\end{equation}
Interestingly, $\lambda_\mathrm{SO}$ is independent of the electron spin momentum in this approximation.  Therefore, we can experimentally extract $\lambda_\mathrm{SO} = v_\mathrm{DQD} / \Omega_\mathrm{L}^\mathrm{D}(k_\mathrm{DQD})$ directly from the measured precession frequency of the spins.  

The Larmor precession frequency of the oscillations shown in Fig.~\ref{excdep} are quite uniform with a frequency $\Omega_\mathrm{L}^\mathrm{D}=0.97$~ns$^{-1}$.  This is similar to the value (1.1~ns$^{-1}$) that we have previously published for comparable DQD acoustic power densities.\cite{Stotz05b}  The slight difference is accounted for by dissimilarities in the mounting of the sample in the cryostat that may have introduced a slightly different static strain of the sample during cooling to 12~K.\cite{Beck06,Crooker05}  Using the value of $\Omega_\mathrm{L}^\mathrm{D}=0.97$~ns$^{-1}$, we obtain a spin-orbit length $\lambda_\mathrm{SO} = 4.2~\mu$m for the 20~nm QW sample, which is expectedly larger than the DQD confinement dimensions $L_\mathrm{DQD}$ of approximately 1~$\mu$m.  As a result, the mesoscopic DQD confinement potential does indeed provide the motional narrowing required to maintain the spin coherence of the microensemble within the DQD.

%
\begin{figure}[tb]
\centering
\includegraphics[width=7cm, clip=true]{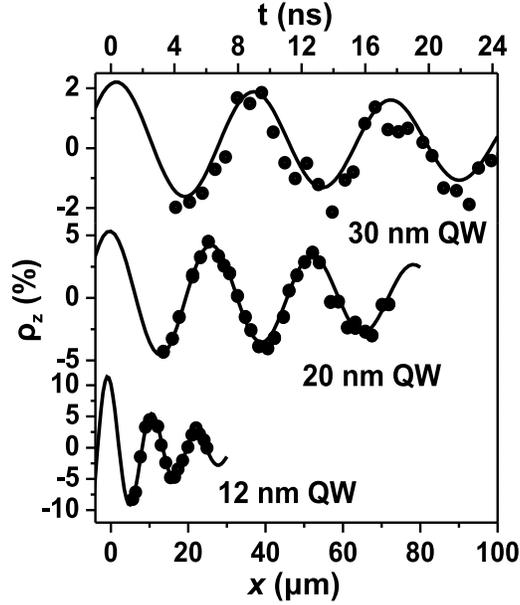}
\caption{Spatial dependence of \pz for QWs with thicknesses of 30, 20, and 12~nm. The symbols and solid lines represent the measured values of \pz and the numerical fits, respectively.  The time axis $t$ is determined by $t=x/\vDQD$.} \label{QWdep}
\end{figure}

\begin{table}
\begin{ruledtabular}
\begin{tabular}{ccccc}
QW & \multicolumn{2}{c}{ $\Omega_\mathrm{L}^\mathrm{D}(k_\mathrm{DQD})$\ \ (ns$^{-1}$) } & $\lambda_\mathrm{SO}$ & $l_s$\\ 
Sample & (Meas.) & (Calc.) & ($\mu$m) & ($\mu$m)\\ 
\hline
30~nm & 0.73 & 0.52 & 5.6 & 200 (194) $\pm$115 \\ 
20~nm & 0.97 & 1.03  & 4.2 & 110$\pm$28  \\ 
12~nm & 2.26 & 2.31  & 1.8 & 17 (20) $\pm$2  \\ 
\end{tabular} 
\end{ruledtabular}
\caption{Spin transport parameters for three different QW samples.  The calculated $\Omega_\mathrm{L}^\mathrm{D}$ uses Eq.~\ref{larmordqd} and a value of $\gamma = 17$~eV\AA$^3$.\cite{Stotz06b}  $\lambda_\mathrm{SO}$ was determined using the measured values of $\Omega_\mathrm{L}^\mathrm{D}$.  The coherence lengths $l_s$ correspond to the fitted curves in Fig.~\ref{QWdep}, and the values in brackets compare the 110~$\mu$m coherence length from the 20~nm QW adjusted by the change in the spin-orbit length $(\lambda_\mathrm{SO})^2$.}
\label{valtable}
\end{table}

The preceding demonstration of the mesoscopic confinement of the electrons spins will now allow us to further explore the relationship between the spin-orbit length, the confinement dimensions, and the coherence length.  According to Eq.~(\ref{larmordqd}), the Larmor precession frequency of the electron spins, and hence the spin-orbit length, can be varied by changing the thickness of the QW. To exploit this dependence, we have performed spin transport measurements on samples with different QW thicknesses.  Figure~\ref{QWdep} compares \pz for the previously discussed 20~nm QW sample with similar samples consisting of single QWs of thicknesses 12 and 30~nm; important parameters from this figure are summarized in Table~\ref{valtable}.  The thinner, 12~nm QW shows a dramatic increase in the Larmor precession frequency that is in good agreement with the value expected using Eq.~(\ref{larmordqd}).  In fact, using the measured values of $\Omega_\mathrm{L}^\mathrm{D}(k_\mathrm{DQD})$ from the 12 and 20~nm QW samples along with the well defined DQD wavevector $k_\mathrm{DQD}$, the spin-orbit parameter $\gamma$ is calculated to be 17 and 16 eV\AA$^3$, respectively, using Eq.~(\ref{larmordqd}).  These are in agreement with our previously determined value of 17$\pm$2 eV\AA$^3$.\cite{Stotz06b}

In our approximation, electron spins at the Fermi surface will experience the same increase in $\Omega_\mathrm{L}^\mathrm{D}(k_\mathrm{F})$ as $\Omega_\mathrm{L}^\mathrm{D}(k_\mathrm{DQD})$ when the QW thickness is reduced.  As shown in Table~\ref{valtable}, this will result in a inversely proportional modification of the spin-orbit length $\lambda_\mathrm{SO}$.  As $\lambda_\mathrm{SO}(\Omega_\mathrm{L}^\mathrm{D})$ is reduced in the 12~nm-thick QW sample, it becomes similar to the spatial dimensions of the DQDs $L_\mathrm{DQD}$.  As a result, the \emph{effective} confinement of the spins is therefore less than that for the 20~nm QW sample leading to shorter coherence lengths $l_s$, which is approximately proportional to the square of the spin-orbit length $(\lambda_\mathrm{SO})^2$.  For the 30~nm-thick QW sample, the measured Larmor precession frequency is larger than that expected using Eq.~(\ref{larmordqd}).  This is attributed to the increasing importance of the strain components to \Bint\ considering the smaller Dresselhaus term for this QW thickness [cf. Eq.~(\ref{larmordqd})] and that the QW is nearer to the surface than in the other samples---the specifics of which will be discussed in detail in a later publication.  Using the experimentally determined $\Omega_\mathrm{L}^\mathrm{D}$, $\lambda_\mathrm{SO}$ is nevertheless determined to be 5.6~$\mu$m.  This larger spin-orbit length is expected to increase the spin coherence length to 194~$\mu$m (given the $(\lambda_\mathrm{SO})^2$ proportionality), and the measured $l_s = 200\pm 115~\mu$m is consistent with this expectation.  The larger error in this measurement is attributed to the fact that the measured transport range is only a small fraction of the long coherence length.  However, the work does indicate that increasing the $\lambda_\mathrm{SO}/L_\mathrm{DQD}$ ratio will enable longer coherence lengths.

Intuitively, the enhanced electron spin lifetimes result from the ability of the mesoscopic confinement potential to rapidly scatter the electron momentum and prevent a spin from undergoing the large precession angles during its mean free path that cause DP dephasing.  Our sample set suggests that the spin coherence length follows a quadratic dependence with respect to the spin-orbit length.  The general relation used to describe DP spin dephasing is\cite{Dyakonov72,Kiselev00}
\begin{equation}
\tau_s \sim [\Omega_L^\mathrm{D}(k_\mathrm{F})]^{-2} \tau_p^{-1} \sim (\lambda_\mathrm{SO})^2 \tau_p^{-1}
\label{DP}
\end{equation}
Equation~\ref{DP} reflects our observed quadratic dependence in $\lambda_\mathrm{SO}(\Omega_\mathrm{L}^\mathrm{D})$ as well as the origin of the long spin coherence times:  rapid momentum scattering $\tau_p$ due to the constant DQD confinement potential.  

The measured thickness dependence of $\tau_s$ for electrons confined by DQDs is, however, quite different than that expected for free electrons in a undoped GaAs QWs.  In the absence of lateral confinement, the spin dephasing will have a similar $[\Omega_L^\mathrm{D}(k_\mathrm{F})]^{-2}$ term associated with the vertical confinement.  The momentum scattering term $\tau_p$, on the other hand, is not dictated by scattering from the lateral confinement potential imposed by the DQDs, but rather by the carrier mobility.  In particular, the electron mobility in GaAs QWs has been shown to vary as $(\mathrm{d}_{QW})^n$, with $n \sim 6$, because of interface roughness scattering,\cite{Sakaki87, Voros05} thus leading to $\tau_p \sim \mu \sim (\mathrm{d}_{QW})^{n}$.  Due to the strong dependence of $\tau_p$ on QW width, the spin relaxation time is expected to decrease with  increasing $\mathrm{d}_{QW}$, in contrast with the experimental results for spin transport via DQDs.

In conclusion, we have shown that the precession frequency, the spin orbit length, and the spin coherence time can be controlled by the QW width.  More importantly, we have demonstrated that the enhanced coherence of electron spins results from the mesoscopic confinement of the DQDs during transport, which does indeed parallel the behaviour observed in stationary quantum dots.  As a result, mobile potentials generated by acoustic fields are anticipated to be a similarly powerful tool in the transport and manipulation of single quantum states within spintronic applications.

We thank J. Rudolph and K.-J. Friedland for comments and for a critical reading of the manuscript, and W. Seidel, S. Krau{\ss}, and M. H\"oricke for their technical support regarding sample fabrication and preparation.  The authors acknowledge the Nanoquit consortium (BMBF, Germany).  J.S. would also like to thank NSERC Canada for financial support.



\end{document}